# Massively-Parallel Implementation of Inextensible Elastic Rods Using Inter-block GPU Synchronization

Przemyslaw Korzeniowski, Niels Hald and Fernando Bello, *Imperial College London*

**Abstract**— An elastic rod is a long and thin body able to sustain large global deformations, even if local strains are small. The Cosserat rod is a non-linear elastic rod with an oriented centreline, which enables modelling of bending, stretching and twisting deformations. It can be used for physically-based computer simulation of threads, wires, ropes, as well as flexible surgical instruments such as catheters, guidewires or sutures. We present a massively-parallel implementation of the original CoRdE model as well as our inextensible variation. By superseding the *CUDA Scalable Programming Model* and using inter-block synchronization, we managed to simulate multiple physics time-steps per single kernel launch utilizing all the GPU's streaming multiprocessors. Under some constraints, this results in nearly constant computation time, regardless of the number of Cosserat elements simulated. When executing 10 time-steps per single kernel launch, our implementation of the original, extensible CoRdE was x40.0 faster. In a number of tests, the GPU implementation of our inextensible CoRdE modification achieved an average speed-up of x15.11 over the corresponding CPU version. Simulating a catheter/guidewire pair (2x512 Cosserat elements) in a cardiovascular application resulted in a 13.5 fold performance boost, enabling for accurate real-time simulation at haptic interactive rates (0.5-1kHz).

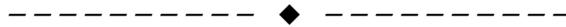

---

## 1 BACKGROUND

One-dimensional deformable bodies - elastic rods - are long and thin bodies able to sustain large global deformations, even if local strains are small. Many approaches have been presented to model them spanning various application fields, most notably, computer graphics (ropes, hairs) and mechanics (wires, cables). In this paper, we focus on computer-based, virtual reality medical simulators. However, the methods presented can be applied to other domains. The last decade has seen growing interest in the benefits of using VR medical simulators in a range of specialties [1]. In addition to accurate simulation, this particular application field requires also real-time interactivity at high rates. This is due to the use of haptic devices – human-computer interfaces, often mimicking real surgical instruments, which track the user's manipulations and generate force feedback. In order for this feedback to be as realistic as possible, it is accepted that the minimum haptic device update rate must be at least 0.5 kHz and preferably 1 kHz [2]. This gives less than 2 milliseconds to get the input from the device, do all the physics computations (i.e. deformable bodies simulation, their interactions with virtual surgical instruments, collision detection and response, integration) and send the resulting feedback back to the user. Processing at this speed is challenging and, so far, it normally involves a trade-off between accuracy and speed.

### 1.1 Related work on elastic rods modeling

The fundamental parts of medical simulators are the underlying mathematical models of organs, tissues and surgical instruments. A realistic, physically-based and real-time model of an elastic rod is a key component for simulators aiming particularly at Minimally Invasive Surgery (MIS), such as endovascular procedures where clinicians insert long, thin, flexible instruments – catheters and guidewires, into the patient's vascular system. Guided by medical imaging (X-Ray Fluoroscopy), the catheter / guidewire pair is navigated into the target diseased vessel (e.g. coronary arteries) to treat the pathology (angioplasty). Another example is gastroscopy or colonoscopy, where clinicians insert, via natural orifices, a flexible endoscope equipped with a camera, light source and often other instrumentarium. Simulators for conventional "rigid" laparoscopy may also require an elastic rod model for modelling sutures or blood vessels.

Various relevant publications on the use of elastic rod models come from two disparate disciplines: hair animation and simulation in interventional radiology. In one of the earliest approaches proposed for hair modelling by Rosenblum *et al.* [3], the rod is discretized as a set of mass-points connected in a chain by linear springs, which try to maintain the given segment length between the neighboring mass-points (mass-spring system - MSS). Additional angular springs behave as hinges, which aim to preserve the given angle between the neighboring segments. Anjyo *et al.* [4] presented an approach based on a simplified cantilever beam. By neglecting a shear deflec-

---

- *P.K., N.H. and F.B. are all with the Department of Surgery and Cancer, Imperial College London*
- *The corresponding author F.B. may be contacted at f.bello@imperial.ac.uk*

tion, they derived simple differential equations that can be solved analytically, therefore preventing hair from stretching. Both solutions did not account for torsional deformations. This was addressed by Selle et al. [5] who added a system of tetrahedral "altitude springs" to a MSS in order to recreate the hair twist, thus enabling the simulation of curly hair. Approaches employing rigid multi-body serial chains connected by either soft springs [6] or hard constraints [7] were also applied to this domain. Further details of the simulation and modelling of hair can be found in a review by Ward et al. [8] and a Siggraph'07 course [9].

In medical simulation, Wang et al. proposed a different approach to recreating material twist in MSS using a scalar torsion angle parameter for interactive suture simulation. [10] and [11] showed the possibility of simulating an elastic rod in real-time and with visually correct accuracy also using a MSS. Thanks to increasing computational power, solutions based on continuum mechanics approaches such as the finite element method (FEM) have become feasible. [12] introduced a non-linear deformable beam model resulting in an accurate simulation, whilst [13] simulates rods, including friction, as a set of straight, non-bendable, incompressible beams with perfect torque control using a quasi-static approach.

The above solutions convincingly reproduce the bending phenomenon of the material, but either ignore the twisting deformation, recreate it in an inaccurate or inefficient way for realistic real-time simulation or handle them in a different way than bend deformations. In [14], Pai introduced an alternative model based on an established theoretical framework – the Cosserat theory of elastic rods [15], which is considered as a final step in the formulation of a modern theory of elastic rods [16]. Pai's model – the Strand – realistically reproduces all the deformations of an elastic rod by considering a continuous oriented curve. Each point lying on a curve has its own local coordinate system - a director. When the curve is undergoing a deformation, the director moves together with the point governing the orientation of the material frame. The strains relate to the spatial difference between these frames. In order to simulate a strand of surgical suture, Pai derived a set of spatial ordinary differential equations (ODEs) that can be efficiently integrated in two passes. However, Pai's solution is limited to static deformations, i.e. it does not explicitly simulate the centerline, but reconstructs it using specified position and orientation of start and end points, which complicates the collision handling [17]. Bertalis et al. [18] modified Pai's approach by using energy minimization to compute the equilibrium position of strands for hair simulation. This allowed for handling external forces such as gravity and collision response, but was still limited to static deformations, thus preventing its use in animation. In his following paper, Bertalis et al. [19] simulated the dynamics of Cosserat rods using Lagrangian mechanics. However, the computation time was quadratic in respect to the number of elements, preventing its application in an interactive simulation of longer rods. In [17], Spillmann and Teschner proposed the CoRdE (French for rope), a model also based on Cosserat theory considering dynamic effects in linear time. The authors then improved the CoRdE model by making it inextensible [20], whilst in [21], Duratti et al. applied a solution resembling the CoRdE model to real-time interventional radiology simulation. Other uses of the CoRdE model include interactive suturing [22]. More recently, Wen et al. solved dynamic deformations of the Cosserat rod using a discrete differential geometry formulation for guidewire and catheter insertion [23].

An interesting approach presented around the same time as the CoRdE model is that by Bergou et al. [24]. They model twist as a deviation from a canonical frame and derive a method to evolve this quasi-statically. Later, Kmoch et al. [25] tailored this model to real-time hair simulation and presented a massively-parallel implementation running on the GPU [26].

Based on the computation times reported in the above publications, real-time interaction at haptic interactive rates (0.5 - 1 kHz) was only possible for rods consisting of a small number of nodes/mass-points. This either limits the length of the simulated rod or its accuracy. Additionally, some specific interactions between the rod and rigid / soft bodies or other rods, necessary for the particular application (e.g. catheter / guidewire pair), can further significantly degrade the performance.

The massively-parallel CUDA implementation presented here is based on a modified version of the CoRdE model with rod inextensibility, collision response, self-collisions and rod(s) interactions handled by our custom iterative constraints framework. It enables accurate real-time simulation of long wires (>3000 Cosserat elements) at haptic interactive rates, fast response to user manipulations and an easy parameterization of the mechanical properties of the rods. By turning off these modifications, the model reverts back to the original, extensible CoRdE model [17]. A similar attempt to accelerate their simplified CoRdE implementation for suture simulation using CUDA was made by Punak et al. [27], but the speed-up achieved was unsatisfactory and they decided to develop their model on the CPU in a serial manner. In [26], Kmoch et al. applied CUDA to their hair animation system based on [24]. but did not report substantial speed gains.

## 1.2 GPGPU on NVidia CUDA

In recent years, graphics processing units (GPUs) have evolved into highly parallel multi-core systems allowing solving general-purpose problems (GPGPU computing). Unlike CPUs, GPUs have an architecture oriented on throughput, specialized in computationally-intensive, highly parallel computation, rather than complex data caching or flow control (**Fig. 1** *Left*). Because the same program is executed for each data element, there is a lower requirement for sophisticated flow control. In addition, since the same program is executed on many data elements and has high arithmetic intensity (the ratio of arithmetic operations to memory operations), latencies due to memory access can be hidden without the need of big data caches.

CUDA (Compute Unified Device Architecture) is a general purpose parallel computing platform and programming model implemented for the NVidia GPUs [28]. It gives developers access to the virtual instruction set and memory of the parallel computational elements in CUDA GPUs. A CUDA enabled GPU is built around a set of Streaming Multiprocessors (SMs). Each SM consists of a number of compute cores (CUDA cores) depending on the chip generation (8-192). Each CUDA core contains an array of integer and floating point arithmetic logic units (ALUs) and special function units (SFUs), which handle specialized instructions such as trigonometric functions, square roots or reciprocals. Each SM is also equipped with a relatively small amount (48-64KB) of fast on-chip memory called shared memory. Shared memory allows threads within the same block (see below) to cooperate on solving a sub-problem. It also enables reuse of data and reduces the traffic to / from off-chip global memory (DRAM, 1-4GB). For many CUDA applications, exploiting the shared memory is the key to achieving high performance gains.

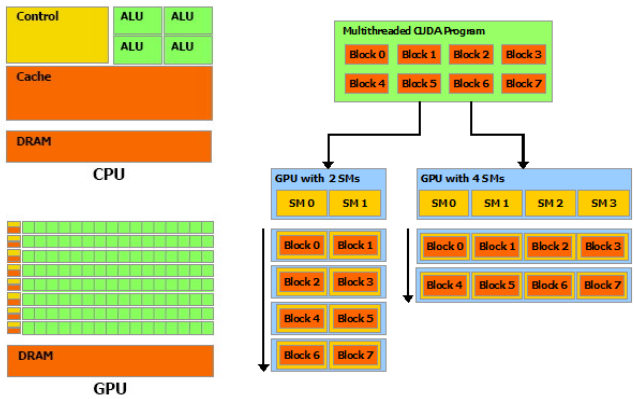

**Fig. 1.** *Left*: CPU vs GPU architecture comparison. *Right*: Scalable Programming Model of CUDA architecture [28]

A multithreaded CUDA program – a kernel – is partitioned into blocks of threads that execute independently from each other. The threads of a block execute concurrently on one SM, and multiple blocks (up to 8 on current generation) can execute concurrently on one SM. As blocks terminate, new blocks are launched on the vacated multiprocessors. In fact, each block can be scheduled on any of the available SM within a GPU, in any order, concurrently or sequentially, so that a CUDA program can execute on any number of multiprocessors located even on different physical GPUs or computing nodes. This decomposition is called *Scalable Programming Model* and is illustrated in **Fig. 1** *Right*. It allows threads to cooperate within the thread block when solving each sub-problem, and at the same time enables automatic scalability. The *Scalable Programming Model* allows for synchronization of all the threads in the same block (intra-block sync) by __syncthreads() function calls. CUDA specification does not define a function for a global, GPU-wide, inter-block synchronization as this would not allow the flexibility in blocks scheduling and, as a result, affect the overall scalability. The only way of doing this supported by the CUDA specification is to finish the execution of one kernel and then launch a new one. However, by accepting some trade-offs, it is possible to supersede the *Scalable Programming Model* and implement inter-block synchronisation. A range of different methods for implementing such synchronisation is presented in [29].

## 2 METHODS

The Cosserat rod is a non-linear elastic rod with an oriented centreline, which enables modelling of bending and twisting deformations. A comprehensive theory is given in the book "*Nonlinear problems of elasticity*" by Antman [15]. Here we summarize the model as explained in [17] and our custom iterative constraints framework, later presenting our massively parallel implementation using CUDA. At the end of the section we also describe our collision and self-collision detection algorithms neccesarry to simulate interactions between the rod and virtual environment, as well as with itself or other rods.

### 2.1 Cosserat Rod

The centreline of the rod is represented by a function mapping line parameter $s$ to a position in 3D space $\mathbf{r} = \mathbf{r}(s): [0, 1] \to \mathbb{R}^3$. To represent the stretching deformation along it we define the strain vector $\mathbf{v} = (v_1, v_2, v_3)^T$, which is a rate of change in the position of the centreline. By assuming that the rod is unshearable $v_1 = v_2 = 0$. The stretch along the centreline is equal to $v_3 = ||\mathbf{r}'||$. An unstretched rod has $v_3 = 1$.

In order to represent bending and twisting deformations, the concept of material frames is introduced. The material frame is an orthonormal basis $\mathbf{d}_k$, $k = 1, 2, 3$, where $\mathbf{d}_k$ are called directors. The first and second director indicate the orientation of the centreline, whereas the third one, $\mathbf{d}_3(s)$, is always adapted to the curve, i.e. parallel to the tangent $\mathbf{r}'(s)$ at the same point (**Fig. 2**). From the directors we derive the rotation matrix $\mathbf{R}(s) \in \mathbb{R}^{3 \times 3}$.

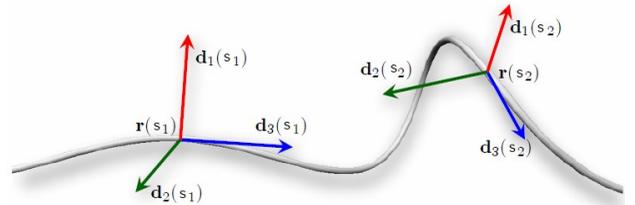

**Fig. 2.** The material frames adapted to the rod's centreline. With permission from [17].

Using differential geometry, from the material frames we obtain an orientational rate of change in the reference frame – the *Darboux vector* – $\mathbf{u}_0 \in \mathbb{R}^3$. Its components are the areas swept by the directors when proceeding from $s$ to $(s + \Delta s)$.

$$u_0(s) = \frac{1}{2} \sum_{k=1}^{3} \mathbf{d}_k(s) \times \mathbf{d}'_k(s) \qquad (1)$$

$\mathbf{d}'_k$ is a partial derivative of the material frame with respect to the line parameter $s$. After rotating $\mathbf{u_0}$ from the reference frame into the local frame $\mathbf{u} = \mathbf{R}^T\mathbf{u_0}$, $\mathbf{u}$ relates to bending and twisting strains (**Fig. 3**).

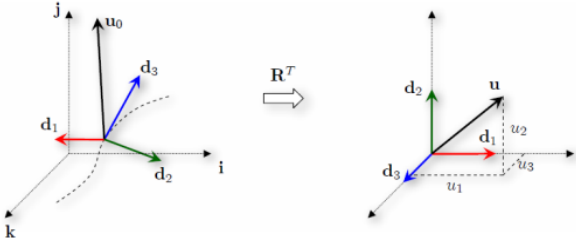

**Fig. 3.** Darboux vector. With permission from [17].

Based on the defined strain rates and assuming a linear strain-stress relationship, we derive the stretch energy of the entire rod:

$$V_s = \frac{1}{2}\int_0^1 K_s(v_3 - 1)^2 ds \quad (2)$$

Where $K_s = E_s \pi r^2$ is a stiffness constant, $E_s$ is a stretching Young's modulus and $r$ is the radius of the rod's cross-section. The bending energy is calculated respectively:

$$V_b = \frac{1}{2}\int_0^1 \sum_{k=1}^{3} K_k(u_k - \hat{u}_k)^2 ds \quad (3)$$

Where $K_1 = K_2 = E_b \frac{\pi r^2}{4}, K_3 = G \frac{\pi r^2}{2}$, $E_b$ and $G$ are Young's and shear modulus governing the bending and torsional resistance, and $\hat{u}_k$ are the intrinsic bend and twist parameters. They are used to control the resting shape of the rod, for example, to model curved tips of guidewires. By minimizing $V = V_s + V_b$ and by treating bending and torsion in a unified manner, we coupled the strain rates together. The twist deformation is balanced out by the bend deformation and vice versa, which can result in the looping phenomenon.

*Discretization*

The centreline of the rod is discretized into $N$ mass (control) points $\mathbf{r}_i \in \mathbb{R}^3, i \in [1, N]$ and $N - 1$ material frames $\mathbf{R}_i \in \mathbb{R}^{3 \times 3}$ as shown in **Fig. 4**:

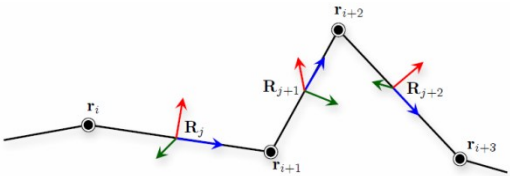

**Fig. 4.** Discretized centreline of the rod with material frames. With permission from [17].

The spatial derivative of the centerline is approximately:

$$\mathbf{r}'_i \approx \frac{\mathbf{r}_{i+1} - \mathbf{r}_i}{\|\mathbf{r}_{i+1} - \mathbf{r}_i\|} \quad (4)$$

In order to provide a singularity-free parameterization, the material frames $\mathbf{R}_j$ are expressed by unit quaternions $\mathbf{q}_j$. Thus, the spatial derivative is approximated as:

$$q'_j = \frac{1}{l_j}(q_{j+1} - q_j) \quad (5)$$

where $l_j$ is the step length. Using quaternions enables to fully determine the material frame without considering the centerline points. The actual strain rates can be computed as:

$$u_k = 2B_k q \cdot q' \quad (6)$$

where $\mathbf{B}_k \in \mathbb{R}^{4 \times 4}$, $k = 1, 2, 3$ are constant skew-symmetric matrices. The detailed derivation of $\mathbf{B}_k$ is given in the appendix of [17]. The CoRdE model also considers internal friction forces which damp relative motion in the rod. They are modelled as additional dissipation energy terms, both translational and angular [17].

*Constraints*

The bending energy as defined above depends solely on the configuration of the material frames. In order to couple the material frames with the centreline, we define a parallel constraint $C_p$, which aligns the third director $d_3$ of the material frame with the tangent $\mathbf{r}'$ of the centreline.

$$C_p = \mathbf{r}' - d_3 = 0 \quad (7)$$

To maintain the above constraint, the penalty method is used resulting in the following penalty energy equation:

$$E_p = \frac{1}{2}\int_0^1 K_p \left(\frac{\mathbf{r}'}{\|\mathbf{r}'\|} - d_3\right) \cdot \left(\frac{\mathbf{r}'}{\|\mathbf{r}'\|} - d_3\right) ds \quad (8)$$

where $K_p$ is a numerical spring constant that depends on the simulated material, and $\|\mathbf{r}'\|$ is the element length. The advantage of the penalty method is its simplicity. It acts as an additional force on the mass-points and torque on material frames and it can be enforced locally. The disadvantage is a possible loss of accuracy, as the constraint may not be always exactly satisfied. In addition, higher values of $K_p$ necessary for the simulation of stiffer materials require smaller time step. In [20], the authors eliminate the penalty method using more accurate, but less efficient Lagrange multipliers.

In many applications having an extensible rod is not desirable. Reducing its stretching by increasing the $K_s$ constant to higher values would also require decreasing the time-step. However, assuming rod inextensibility, we can omit calculating $V_s$ and replace it with a distance constraint $C_d$, which will try to maintain the desired rest length between the centreline's mass-points. To do this we employ Lagrange multipliers using the $\mathbf{JM}^{-1}\mathbf{J}^T$ projection. Additionally, we employed constraints to model other aspects of the simulation. Binding constraints $C_b$, which are effectively distance constraints with a desired rest length set to zero, keep the two rods together. Contact constraints $C_c$ are responsible for handling collisions and self-collisions with Coulombian friction.

By solving constraints globally, i.e. all at once in a matrix form as suggested in [20], we loose the possibility to parallelize the model. Therefore, we apply distance, collision and binding constraints locally using iterative impulse-based methods instead. Such a way of solving constraints is often used in modern physics engines such as Bullet or NVidia PhysX. In other words, to preserve the constraint, we compute a sudden change in momentum – an impulse – which, input into the equations of motion, eliminates velocities that violate the constraints. We can call these constraints "hard constraints", as opposed to "soft constraints" enforced using the penalty method at the force (acceleration) level in case of parallel constraints.

In sum, after a force is applied to the rod by a user manipulating it, for example with a haptic device, we differentiate the bending energy $V_b$, if needed, the stretch energy $V_s$, and the parallel constraint penalty energy $E_p$ with respect to the coordinates to obtain the stresses, i.e. restitution forces and torques which accelerate the centre-line mass-points and material frames to equilibrium. Having the mass-points $r_i$ loosely coupled by the parallel constraints with the quaternions $q_j$ allows for their independent force integration. The mass-points are integrated as if they were particles and the quaternions as if they were representing orientations of rigid bodies. After force integration, we constrain the resulting velocities using impulses and integrate them into new positions and orientations using a semi-implicit Euler scheme. As all the calculations (including constraints) are performed locally, the CoRdE model is a suitable target for GPU acceleration.

## 2.2 CUDA Implementation

The multi-thread implementation consists of three main threads: physics simulation (Coserat rod, collision detection), graphics and input (haptics).

The physics simulation can be toggled between the CPU and the GPU. During each GPU update, a single CUDA kernel is invoked taking as parameters the number of time-steps to process, the number of iterations over the constraints, material properties and input from the user (haptic device position vector). The graphics thread runs at standard 60Hz in the CPU. During each update, it requests a copy of the Cosserat rod mass-points positions to be transferred from the GPU to CPU for rendering purposes. The thread handling communication with the input (haptic) device runs at 0.5-1 kHz and can request the state of the force-feedback vector, as well as provide a new position vector at any time from the physics simulation thread. This is possible because CUDA GPUs are equipped with separate data copy engines independent of kernel execution. This allows for fast, synchronous or asynchronous transfers (using pinned memory and different CUDA stream) of the intermediate data between the global memories of the GPU and CPU without waiting for the kernel to finish processing the specified number of physics simulation iterations. This enables writing the current state of the haptic device to the running kernel and getting back the force-feedback value at the rate higher than that of actual kernel executions.

All the data arrays containing the initial positions of mass-points and quaternions, velocities and external forces are copied to the global memory of the GPU at the application initialization. During each kernel launch, i.e. physics simulation update, these arrays are loaded from the global memory (DRAM) to the fast shared memory of the SM in order not to duplicate costly global memory operations. Each thread processes a single Cosserat element, i.e. a mass-point and a corresponding quaternion. To achieve this, it requires access to information about the position and velocity of the next neighbouring element. Processing a single time-step per kernel launch (1ts/kl) on the GPU can already yield a performance boost. However, the available arithmetic power is largely counterbalanced by the overhead of launching a kernel and global memory access latencies.

Larger speed-ups are achieved by running multiple time-steps per kernel launch ($n$ ts/kl). In a single-block kernel implementation, a specific case when all the mentioned arrays fit into the shared memory of a single SM can be done within the boundaries of the *Scalable Programming Model*. By careful intra-block synchronization between the threads using the __*syncthreads()* CUDA keyword, we can keep the data in the shared memory between the subsequent time-step iterations. Additionally, calling __*syncthreads()* ensures shared memory consistency, i.e. changes made by one thread are visible to all the other threads within the same block. Running multiple iterations over the constraints requires additional position and velocity information exchange between the neighbouring mass-points, and thus two extra __*syncthreads()* calls per constraints iteration.

In this single-block case, small (48-64KB per block) shared memory size limits the number of the Cosserat elements which may be processed to around 512, which in turn limits the maximum length or accuracy of the simulated rod. As the current generation of GPUs can handle many more threads per block (currently up to 2048), it may be possible to fit more Cosserat elements into the block by sacrificing some performance and moving data back to the global memory. However, considering handling collision detection during each time-step, which requires each thread to maintain its own stack in the shared memory for efficient iterative AABB tree traversal, in our implementation a single block is able to handle only around 256 mass-points maximum.

Utilizing just a single block occupying a single SM on a GPU that currently can have up to 15 of them (e.g. NVidia GeForce GTX780Ti) is clearly not an efficient use of resources. As mentioned earlier, the *Scalable Programming Model* does not allow for global, inter-block synchronization and communication. In case of a multi-block kernel, such functionality is essential to correctly simulate even a single time-step per kernel launch with Cosserat elements spanned along several SMs. In case of an extensible rod, when using penalty forces rather than iterative distance constraints, this problem may be solved by splitting the kernel into two smaller ones as the only possible way of GPU-wide syncing is by launching a new kernel. The first kernel would handle only the calculations of elastic forces. The second one would handle forces and velocities integration. This obviously adds launch overhead and, more importantly, prevents from keeping the data in the shared memory as it is discarded at each kernel completion.

However, by superseding the *Scalable Programming Model* and accepting some trade-offs in portability across platforms / GPUs, it is possible to utilize all of the SMs available on the GPU. Using an atomic counter held in the global memory, a *__syncgpu()* method, which works as a barrier blocking the complete thread block until other blocks reach the same point [29], may be implemented. Of course, in this case we cannot schedule more blocks than there are SMs on the GPU, as this would result in some blocks never being processed, and others waiting indefinitely for them to reach the same point. The GPU would stall often resulting in a system crash.

Another problem is the inter-block or rather inter-SM communication. Data held in shared memory is only visible by threads residing in the same block. The only way to communicate between the blocks running on different SMs is via global memory. Given that in the Cosserat rod simulation there is mainly a need to exchange the data between the last and the first mass-point/quaternion of the bordering blocks, the performance penalty is not that substantial. To make this work, the pointers referring to global memory data must be marked as *volatile*. This ensures that all the memory accesses will result in an actual memory read/write operation, i.e. data will not be cached in registers or L1 cache.

```
load the data from the global to shared memory
for i physics iterations (time-steps)
{
        detect collisions
        apply external forces and torques
        calculate Cosserat elastic forces and torques
        integrate forces and torques
        for j constraint iterations
        {
                apply distance constraints impulses
                apply collision response impulse
                if(selfCollisions)
                        apply self-collisions impulse
                if(rodsBinding)
                        apply binding impulses
        }
        integrate linear and angular velocity
}
write the data from shared to global memory
```

**Fig. 5.** The algorithm of the multi-block, inextensible Cosserat rod kernel. The *lines in italic* indicate the functions which require data exchange with other Cosserat element(s) often residing in a different thread block/SM.

This solution enables utilization of all the SMs and significantly increases the possible length of the simulated instrument without affecting the performance. In other words, as long as our GPU can accommodate the desired number of Cosserat elements, the computation time is nearly constant, regardless of the length of the simulated rod. In **Fig. 5** we summarize the algorithm used, highlighting the functions requiring data exchange with other Cosserat element often residing in a different thread block/SM.

## Collisions and self-collisions

We implemented a collision detection scheme running serially on the CPU or parallely on the GPU. An axis-aligned bounding box (AABB) hierarchy guides the broad-phase collision detection stage, and a brute-force approach is used for the narrow-phase. Due to the hollow shape of the 3D models of human anatomy used in our application field (endovascular interventions, flexible endoscopy), the collision detection algorithm uses volume partitioning rather than the more common spatial partitioning such as BSP or OCT trees. This results in a well-balanced tree, which is more stable to traverse in terms of performance. The AABB tree is pre-computed on the CPU and the resulting arrays containing the mesh triangles and the tree structure are copied to the GPU global memory at the initialization stage. During the broad-phase, the tree is searched iteratively using a small stack residing in the shared memory. Its size depends on the maximum depth of the tree, which is usually between 14 and 18 levels for complex anatomical models. In order to reduce thread divergence, the broad-phase reports the indices of intersecting bounding boxes and stores them in another small stack. In the narrow-phase, we perform a brute-force collision check against the triangle(s) in the reported bounding boxes. Next, a collision response vector is calculated as a weighted average of all normal vectors of colliding triangles. The weight depends on the penetration depth for the given triangle. In this paper we use sphere-triangles collisions, but this scheme can support box-triangles and ray-triangles as well. The GPU collision detection can be used on its own, launched as a separate kernel, or invoked from within the simulation loop running on the GPU.

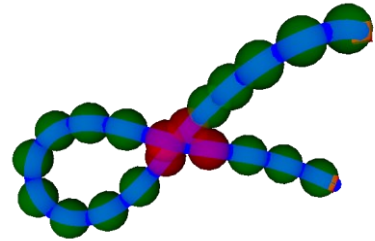

**Fig. 6.** Self-collision detection: The mass-points of the centerline are colored in blue. Larger green and red spheres are respectively non-colliding and colliding broad-phase bounding spheres.

In the case of self-collisions or collisions with another rod, we developed a broad-phase stage based on bounding spheres wrapping a number of neighbouring mass-points (**Fig. 6**). First, each bounding-sphere is checked against all others. Then, the colliding pairs are reported for a narrow-phase sphere-sphere check. If mass-points overlap, the collision response vector is stored to be used at the constraints solving stage.

## 3 RESULTS AND DISCUSSION

Below we report on the performance (computational times in ms) achieved on a desktop and a laptop computer with mobile GPU chipset. The desktop was a HP x4600 workstation (Intel Core2 Quad @2.66 GHz, 8GB RAM) with NVidia GeForce GTX 560 (7 SMs @1.66GHz, 336 CUDA cores total, 1GB DRAM) running 64bit Linux (Ubuntu 12.10). The laptop was an ASUS N55SL laptop (referred onwards as ASUS, Intel Core i7 @2.2 GHz, 6GB RAM) with NVidia GeForce GT 635M (3SMs @1.35GHz, 144 CUDA cores total, 2GB DRAM) running Win7 x64. The testing environment was developed in Java7 using the *JCuda* wrapper to communicate with CUDA C and the *JME3* graphics engine for visualization. All the tests use our inextensible CoRdE modification and constraints framework explained in section 2.1.

It is worth noting that the HP desktop was equipped with a single GPU that had to switch the context between OpenGL rendering and CUDA computing, which slightly slowed down the computations. This was not the case of the ASUS laptop, which was equipped with two separate GPUs (*NVidia Optimus*™ technology), therefore the NVidia GeForce GT635m was entirely dedicated to CUDA calculations. On the other hand, the Windows Driver Model (WDM) on Microsoft platforms adds around 0.1ms delay to each kernel launch and memory copy operations. This is why we decided to run Linux on the HP testing set-up.

A total of five different tests were conducted. The first test (free space) demonstrates pure performance aspects of the Cosserat rod elastic forces and constraints framework calculations. We compare our results to other CoRdE implementations including the original [17] and inxtensible [20]. Secondly, we briefly present the performance of our parallel AABB collision detection algorithm. In the third test, we analyze in detail a combination of the Cosserat rod and collision detection in a real-life example – a gradual rod insertion into a 3D anatomical vascular model. Additionally, we compare the results between desktop and mobile platforms. The fourth test extends the previous one by adding another rod of the same length and interactions between them in a catheter/guidewire manner, i.e. one rod is inserted into another and can move forwards and backwards within it. In the fifth and final test, we check the performance of the self-collisions by tying two rods together using a double Fisherman's knot. The speed-up in performance achieved in all the tests is then summarized and discussed.

The numbers in square brackets [] in the following charts indicate the number of physics time-steps (iterations) simulated per single physics update/kernel invocation. On the CPU this may not have much meaning as the computation time scales linearly, but for the GPU this yields significant performance gains as described in the following paragraphs.

### 3.1 Free space test

This test focused solely on the Cosserat rod elastic forces, the distance constraints and integration performance. The rod was locked at one end and released like a pendulum, swaying for a few seconds in free space. We measured the performance of 1, 10 and 20 physics time-steps simulated per single kernel launch/physics update. Each time-step internally ran 10 iterations over the distance constraints. No collision detection or self-collisions were involved. The maximum number of mass-points able to fit in our GPU (7 SMs) was 6x512=3072. One SM was intentionally left free for graphics rendering. For visualization purposes, calculation times were extrapolated up to 4092 mass-points. All the times given below for the GPU implementation are total execution times "visible" from the host (CPU) side. They include kernel launch overhead, as well as transferring position data of the mass-points back to the CPUs main memory for further processing and/or rendering.

*GPU implementation performance characteristics*

**Fig. 7** presents the behavior of the GPU implementation for 1, 10 and 20 time-steps simulated per single kernel launch. For a single iteration, a 16 fold increase in number of Cosserat elements (0.1742ms for 128 vs. 0.1942ms for 2048) results in just over an 11% increase in computation times. For higher number of iterations, this figure grows to 25% for 10 and 27% for 20 iterations. This small, yet noticeable increase is caused mainly by the inter-block synchronization mechanisms and, to a smaller extent, by global memory reads/writes to communicate between the SMs. In **Fig. 7** we can also see that increasing the number of physics time-steps from 1 to 10 and to 20 per kernel launch, yields an average increase in computation times of only x2.72 and x4.50 respectively. In other words, the overall average cost of a single time-step when executed on the GPU in a batch of 10 per kernel launch was 73.2% (0.49ms/10) lower than executing a single time-step per kernel launch (0.18ms/1). In a batch of 20, the overall average cost is further reduced to 77.5% (0.81ms/20). This is caused by the fact that, after overcoming the initial overhead of passing the control to the GPU and latency issues related with accessing the data in the GPU global memory (DRAM), the cost of subsequent physics iterations is significantly lower. The initial overhead is promptly compensated by the high arithmetic power of the GPU.

The overall average cost of a single time-step simulation when time-steps are executed on the GPU in a batch is a valid performance metric. This is because CUDA GPUs are equipped with separate and independent data copy engines. This allows for fast, asynchronous transfers (using pinned memory and different CUDA streams) of the intermediate data between the global memories of the GPU and the CPU without waiting for the kernel to finish processing the given number of iterations. This, for example, enables writing the current state of the haptic device to the running kernel and getting back the force-feedback value at the rate higher than the rate of actual kernel executions.

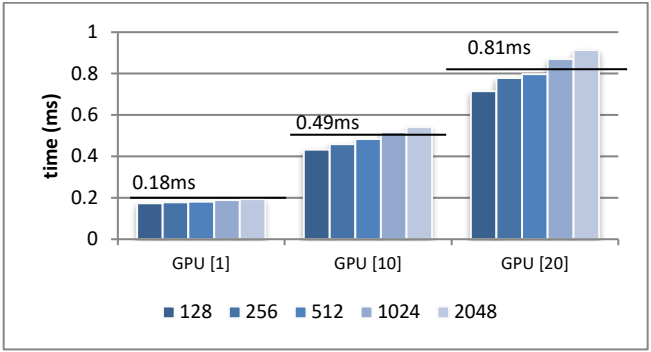

**Fig. 7.** Comparison of number of iterations simulated on the GPU on the HP desktop.

The above properties enable increasing the number of simulated Cosserat elements and, at the same time, improving the accuracy of the simulation, by decreasing the time-step, at very low additional computational cost. For the CPU version, both of these metrics behaved linearly as shown in the next section.

*Comparing GPU vs. CPU performance*

The CUDA version was faster than an almost identical CPU implementation for all the number of mass-points simulated (**Fig. 8**). When running a single time-step per physics update, it was x2.25 times faster even for the smallest number of mass-points (128). It achieved a speed-up of one order of magnitude for approximately 768 mass-points, peaking at x42.4 speed-up for 3072 mass-points.

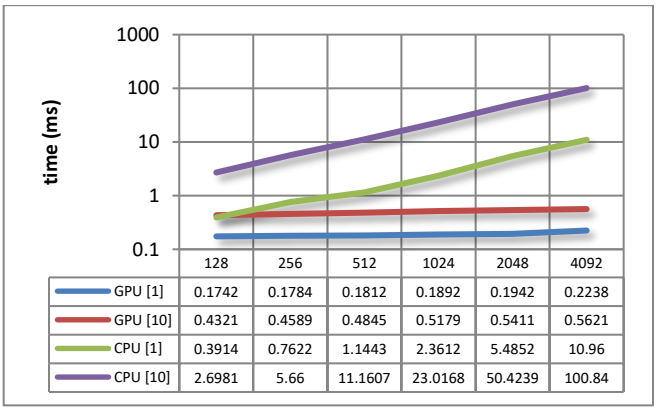

**Fig. 8.** Performance of the Cosserat rod suspended in free space without collisions. Note the logarithmic scale on the vertical axis.

Increasing the number of iterations simulated per physics update from 1 to 10 significantly improved the results. The performance boost reached an order of magnitude (x12.3) for only 256 elements, two orders of magnitude (x100) for around 2200 elements, and x149.0 speed-up for 3072 mass-points.

*Single-block vs. multi-block kernel*

The above results were achieved using the most universal kernel implementation supporting inter-block synchronization. However, if the given number of Cosserat elements is small enough to fit into a single thread block, as described earlier, we can discard the inter-block synchronization and communication overhead between the subsequent iterations, thus reducing calculation times even further by 35-40% (**Fig. 9**). This results in speed-ups over the CPU of x10.4 (0.26ms), x19.5 (0.29ms) and x34.9 (0.32ms) for 128, 256 and 512 mass-points, respectively, compared to x6.2, x12.3, x23.0 speed-ups achieved by the universal kernel. The limit of 512 elements still allows for simulation of relatively long rods and the remaining blocks may be employed for other tasks.

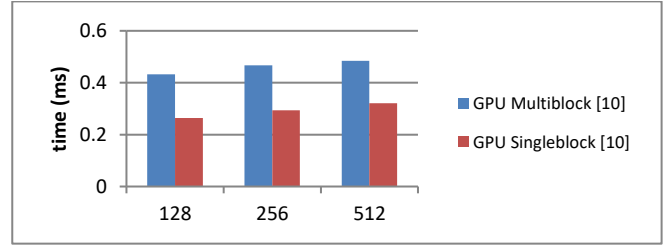

**Fig. 9.** Performance of a universal, multi-block kernel vs. single-block kernel running 10 time-steps per kernel invocations

*Comparing our implementation with others*

In **Table 1** we compare the performance of our extensible and inextensible Cosserat rod implementations to others found in the literature. We chose our inextensible, multi-block GPU accelerated version as the reference. In the case of running 10 iterations per kernel launch (10 ts/kl), we used the overall average cost of a single time-step as the performance metric, i.e. dividing the total computational time by the number of iterations taken. The time measures were taken for 100 and 1000 elements running elastic force calculations, constraints (if any) and time integration, which was analogous to the previous free-space test. Both our extensible and inextensible serial CPU implementations were slower than the original CoRdE [17], and significantly slower than the simplified CoRdE [27].

However, for 1000 Cosserat elements, our reference implementation outperformed the original CoRdE by a factor of x23.85 and the simplified CoRdE by an order of magnitude (x10.5) running 10 time-steps per kernel launch, and by x6.56 and by x2.88, respectively, running 1 ts/kl. Our extensible GPU CoRdE implementation was x1.68 faster than our inextensible (reference one) and x40.0 faster than the original CoRde [17] (10ts/kl). Running a single iteration per kernel launch still yielded a decent speed-up of x5.99 of our corresponding extensible CoRdE implementation over the original one.

Comparing our reference model to the original inextensible CoRdE [20] yields even better results (x300 for 1ts/kl and >x1000 for 10ts/kl). However, we need to take into account that the times in [20] for 100 and 1000 elements were not given explicitly. We derived them from the coil embolization example where authors state constituent times of a simulation of 40 Cosserat elements: 1.9ms for elastic forces, 0.3ms for constrained Lagrange mechanics and 0.06ms for numerical integration. Sum-ming these (2.26ms) and linearly extrapolating to 100 and 1000 elements gives approximated times of 5.65ms and 56.5ms respectively. We did not take into account 8.25ms of collision detection. Also worth noting that wrong units of microseconds instead of milliseconds were given in [20]. Moreover, the model proposed in this paper does not use the penalty method. Longer computation times are compensated by higher stability and larger time-steps.

**Table 1.** Performance of our extensible and inextensible Cosserat Rod GPU and CPU implementations vs. others for 100 and 1000 Cosserat elements in miliseconds.

| Implementation | PC CPU | Inex-tensible | 100 El. (ms) | 1000 El. (ms) |
|---|---|---|---|---|
| Our GPU, 10 TS/KL[1] | Core2 (HP) 2.66 GHz | yes | 0.43/10 | 0.52/10 |
| Our GPU, 1 TS/KL[1] | Core2 (HP) 2.66 GHz | yes | 0.174 | 0.189 |
| Our GPU, 10 TS/KL[2] | Core2 (HP) 2.66 GHz | no | 0.26/10 | 0.31/10 |
| Our GPU, 1 TS/KL[2] | Core2 (HP) 2.66 GHz | no | 0.173 | 0.207 |
| Our CPU, 1 TS/KL | Core2 (HP) 2.66 GHz | yes | 0.391 | 2.361 |
| Our CPU, 1 TS/KL[2] | Core2 (HP) 2.66 GHz | no | 0.160 | 1.436 |
| Original CoRdE [17] | Xeon 3.8 GHz | no | 0.131 | 1.240 |
| Inextensible CoRdE [20] | Core2 3.0 GHz | yes | 5.65* | 56.5* |
| Simplified CoRdE [27] | Core i7 2.93 GHz | no | 0.059 | 0.545 |
| Wen Tang [23] | Core2 2.8 GHz | yes | 0.618* | 8.72* |

[1] our reference inextensible GPU implementation
[2] our equivalent implementation of the original CoRdE [17]
* approximated times as explained in the text

A comparison to a non-CoRdE based Cosserat rod implementation [23] also yields a significant speed-up of x46 for 1ts/kl and x168 for 10ts/kl. Again, the authors only give the simulation time for a 900mm long rod (7.85ms) without stating the underlying discretization. Thus, we made an assumption that the authors used 1 Cosserat element per 1mm of rod's length and linearly extrapolated to 1000 elements to obtain 8.72ms.

### 3.2 Collision detection

This test focused on the performace of the collision detection exclusively. No physics was simulated on the GPU. The rod was inserted into a 3D reconstruction of a vascular silicon phantom model consisting of 26066 triangles represented by an AABB tree of maximum depth of 15, consisting of 46205 nodes and 23103 leaves (**Fig. 10**).

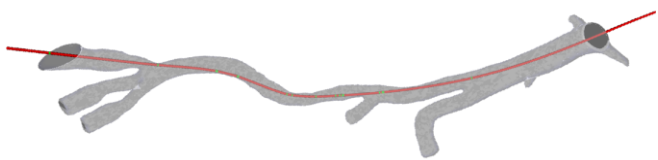

**Fig. 10.** Cosserat rod consisting of 512 elements inserted into a 3D reconstruction of a vascular silicon phantom model.

On average, there were 1.12 triangles per leaf. We used a collision detection kernel described earlier. Before launching the kernel, an array containing the positions of all the mass-points had to be transferred to the GPU memory and, after the kernel execution, an array of collision response vectors was copied back to the CPU memory, which added a substantial overhead as shown in **Fig. 11**.

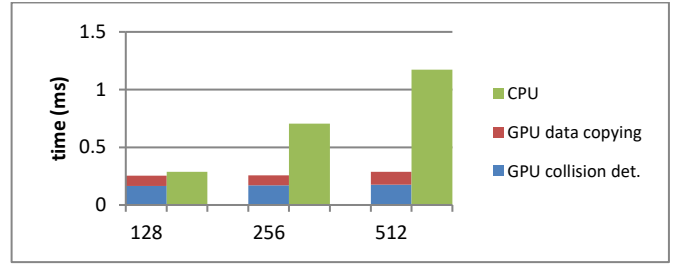

**Fig. 11.** Collision detection performance comparison

**Fig. 11** illustrates that there was not much gain from using the GPU for 128 mass points. However, as the CPU computing times grew linearly with the number of mass-points, the GPU suffered almost no penalty. For 512 points, it was x4 times faster than the identical algorithm running on the CPU (0.29ms vs. 1.17ms). The 3D model used in the simulation was not long enough to fully accommodate rods longer than 512 elements. In a longer model of a complete aorta (18k triangles, 36k nodes, 18k leaves) with 3072 mass-points colliding (6 thread blocks, 512 mass-points each), performance on the GPU was more than x20 times faster than on the CPU.

### 3.3 Guidewire insertion test

A single instrument consisting of 512 Cosserat elements was gradually inserted into the reconstructed 3D phantom vascular model, same as described above. Such number of elements is sufficient to simulate with high accuracy a guidewire navigating from the femoral artery to the heart vessels, or a flexible endoscope inside a colon. The rod did not self-collide. Each kernel invocation processed 10 time-steps, each step containing a broad and narrow phase of collision detection, identical to the one described in the previous section, and 10 iterations over distance and collision constraints including Coulombian friction. In **Fig. 12** we can see that the computation times largely depended on the instrument insertion depth due to the increased number of AABB tree queries. This held true for both CPU and GPU and resulted in a performance drop of up to a half when compared to zero insertion depth.

The average computation time throughout the insertion procedure for CUDA implementation on the HP desktop running 10 iterations per kernel launch was 1.88ms (min. 0.99ms, max. 2.11ms) and was, on average, x8.75 times faster than the CPU version (avg. 16.45ms, min. 11.37ms, max. 20.15ms). A single time-step per kernel launch running on the GPU (0.29ms) was still x5.65 times faster than the CPU (1.64ms). The overall cost of a single GPU time-step, while executed in a batch of 10 per kernel launch (1.88ms/10), was 35% cheaper than launching a single physics iteration.

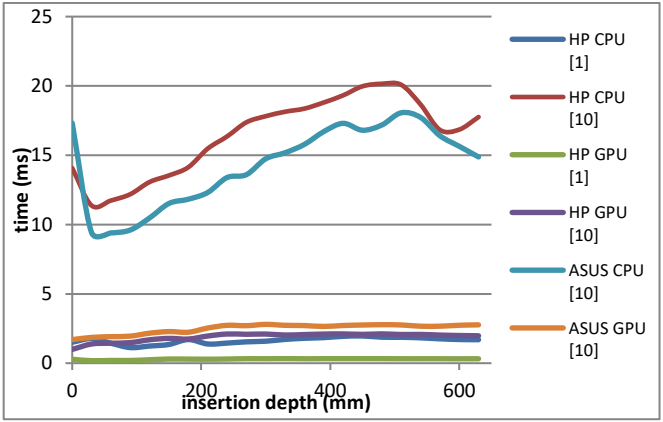

**Fig. 12.** Calculation time at different insertion depths for 512 mass-points, 10 iterations per physics update/kernel launch

**Fig. 12** also shows the performance of the ASUS platform. It achieved only a x5.75 speed-up since the laptop has a much slower GPU (3 SMs vs. 7 SMs on the HP, lower GPU clock rate), but newer and faster CPU. Nevertheless, the GPU implementation on the laptop was running only 25% slower than on the desktop (10 iterations).

The collision detection clearly was a performance bottleneck for both CPU and GPU. Its impact on the GPU performance, especially of the broad-phase, can be minimized by, among others, reducing the thread divergence during AABB tree searchers. Also, various "hacks" can be applied, for example, slightly increasing the AABBs sizes and executing the broad phase only once per kernel launch reduced the average time by more than half, from 1.88ms to 0.81ms.

### 3.4 Catheter and guidewire insertion

This test is similar to the previous one, but with two rods inserted interacting with each other (**Fig. 13**). Each rod consisted of 512 Cosserat elements. As before, each kernel invocation processed 1 or 10 time-steps per physics update, each step containing broad and narrow phases of collision detection and 10 iterations over distance and collision constraints.

We implemented two variants of instruments interaction. In variant one (v1), the instruments interaction was one-way only, i.e. one of the instruments was fully dependent on the other one. This can be illustrated, for example, by a soft guidewire inserted into a stiff catheter. The guidewire's influence on the overall catheter-guidewire pair dynamics is negligible. In a slower but more realistic variant two (v2), the interaction between the two rods was bidirectional with each instrument influencing the dynamics of the catheter/guidewire pair according to its mechanical properties. It is worth noting that the difference between these two variants was quite significant for the GPU version, but practically irrelevant in the CPU implementation. This is because there is more intense synchronization and communications between the blocks in a constraints loop. In case of the binding constraints, especially for v2, the algorithm needs to exchange data about all the Cosserat elements of the other rods via slow global memory. For comparison purposes, the performance of a variant zero (v0) that involved no interactions between the instruments was also considered. The rods were independent and did not collide with each other.

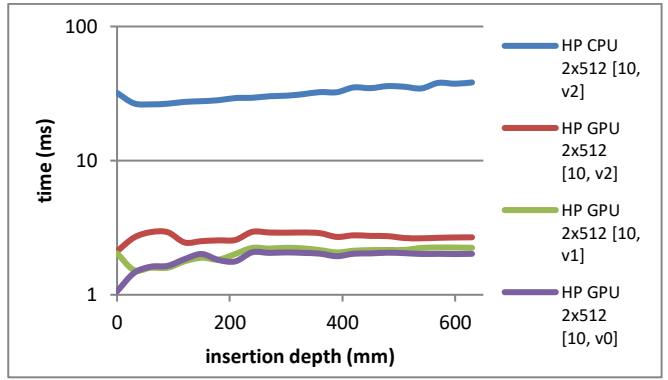

**Fig. 13.** Insertion performance of two instruments

The average computation times for the CUDA implementation on the HP desktop were 1.96ms, 2.15ms and 2.80ms for variants 0, 1 and 2 respectively. The interactions (binding constraints) added 10% (v1) and 43% (v2) overhead for the given number of elements (2x512) compared to practically no overhead on the CPU. Nevertheless, the GPU implementation was x19.3 (v0), x17.6 (v1) and x13.5 (v2) times faster than variant 2 running on the CPU (37.88ms).

**Fig. 14** presents variant 1 and 2 running a single time-step per kernel launch. In this case, the performance difference between these two running on the GPU was negligible (v1 - 0.44ms, v2 - 0.46ms). On average, they were x8.0 times faster on the GPU than on the CPU (3.68ms). The overall cost of a single time-step while executing a batch of 10 iterations per kernel launch (2.8ms/10) was nearly 40% less than launching a single physics iteration (v2).

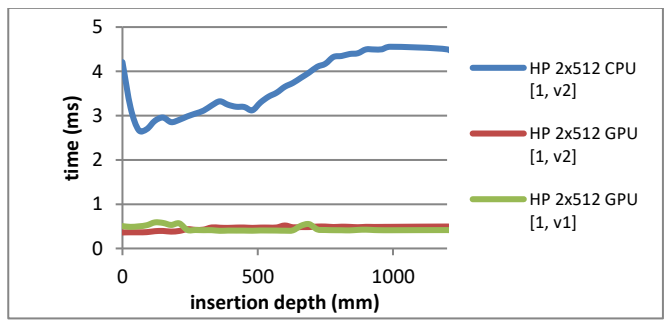

**Fig. 14.** Two instrument insertion test

### 3.5 Double Fisherman's knot tying test

**Fig. 15** shows two stages of tying a double Fisherman's knot. Two Cosserat rods consisting of 256 elements each were attached at different points in space at the ends. We used two Phantom Omni haptic (www.sensable.com) devices to control the loose ends and manually tie the knot. The average number of collisions for the tied knot was 300 colliding sphere-sphere pairs. We used a broad-phase based on bounding spheres (**Fig. 6**) as described in the section on collision detection. The total calculation time on the GPU was 0.32ms for a single iteration per kernel launch and 1.72ms for 10 iterations. On the CPU these numbers were 2.67ms and 21.99ms, giving speed-ups of x8.34 and x12.78, respectively.

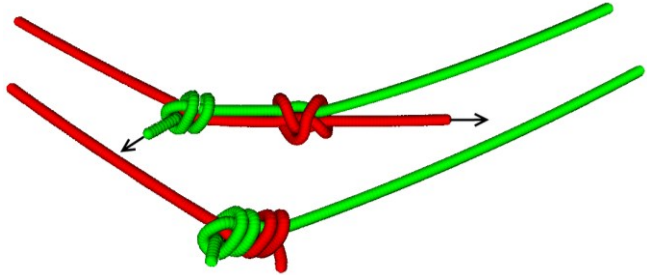

**Fig. 15.** Tying a double Fisherman's knot

We chose the double Fisherman's knot since it has been previously used in a paper on knots simulation using the CoRdE model [30]. Although our overall approach to knot simulation is different, our implementation is x41.5 faster on the GPU (0.32ms compared to 13.3.ms) and x5.0 faster on the CPU (2.67ms compared to 13.3.ms). Even comparing to the fastest adaptive rod version reported in [30] (7.3ms), our GPU implementation still attains a 20 times speed-up. This was in spite of having to use twice as many Cosserat elements to be able to tie this knot and keep it stable, due to our self-collision detection based on spheres rather than actual centerline geometry, which required increasing the number of spheres to make them densely overlap each other (**Fig. 15**).

Comparing our results to the suture simulation based on a fast, simplified CoRdE by Punak et al. [27], our inextensible GPU solution was x7.84 faster than their serial version (0.32ms compared to 2.51ms) and x17.9 faster than their CUDA attempt (0.32ms compared to 5.73ms) for the same number of Cosserat elements (512).

## 4 CONCLUSION

We present a CUDA-based massively-parallel implementation of a Cosserat rod simulation framework and its modification to ensure inextensibility. By superseding the *CUDA Scalable Programming Model* and using inter-block synchronization, we managed to simulate multiple physics time-steps per single kernel launch utilizing all the GPU streaming multiprocessors. Under some constraints, this results in nearly constant computation time, regardless of the number of Cosserat elements simulated. Moreover, improving the simulation accuracy by decreasing the time-step size and increasing their number, results in relatively low additional computational cost.

Comparing our results to other Cosserat rod implementations, when running 10 iterations per single CUDA kernel launch, we achieved speed-ups of at least an order of magnitude when simulating 1000 Cosserat elements on a consumer level GPU. Our extensible GPU CoRdE implementation was x40.0 faster than the original CoRdE version.

In a series of tests, our inextensible Cosserat rod achieved an average speed-up of x15.11 running 10 time-steps per kernel launch, and an average speed-up of x7.32 running a single time-step per kernel launch over our corresponding CPU version (**Fig. 16**) for a moderate number of Cosserat elements (512-1024).

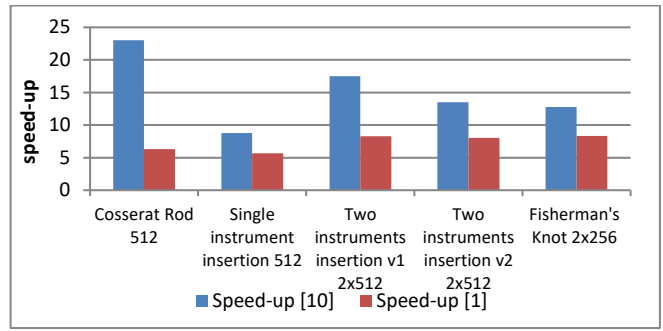

**Fig. 16.** Speed-ups of our GPU implementation achieved over corresponding CPU version in different test sets running 1 and 10 time-steps per single kernel launch.

The first test, a free-space test, showed an interesting performance characteristic of our GPU implementation, namely that of nearly constant computations times. Adding collision detection allowed for more real-life test scenarios such as simulating a guidewire or guidewire/catheter pair navigated in a vascular system. In the last test, we added self-collisions and demonstrated that our solution can also be used to efficiently simulate sutures by manually tying a double Fisherman's knot

The speed-ups achieved enabled for accurate real-time simulation at haptic interactive rates (0.5-1kHz), even when executing a single physics time-step per kernel launch. This makes the presented GPU accelerated model an interesting choice for fast and accurate elastic rod simulation, not only for medical applications, whilst demonstrating that, for some applications, the GPU can deliver a significant performance boost, even when the problem size is relatively small (128-3072 Cosserat elements) in terms of traditional massively parallel computation standards.

### ACKNOWLEDGMENT

This work was supported in part by grants from the EPSRC, the UK Engineering and Physical Science Research Council, the RCUK Digital Economy Programme, the London Deanery STeLI initiative and Health Education North West London.

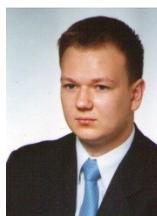

**Przemyslaw Korzeniowski** completed his BSc in Computing at the Polish-Japanese Institute of Information Technology (PJWSTK) in 2009. In 2010 he received an MSc in Computing from Imperial College London. Currently, he is a researcher and PhD candidate at the Department of Surgery and Cancer, Imperial College London, where he is developing virtual reality medical simulators for surgical training. His current research focuses on modelling flexible surgical instruments for endovascular interventions and Natural Orifice Transluminal Endoscopic Surgery (NOTES).

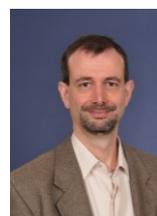

**Fernando Bello** is a computer scientist and engineer working at the intersection of medicine, education and technology. He leads the Simulation and Modelling in Medicine and Surgery (SiMMS) research group, a multi-disciplinary group aiming at building suitable models and simulations of clinical processes, including clinical examination, clinical diagnosis, interventional procedures and care pathways. He is particularly interested in the use of virtual environments and haptics in the context of education and training. Bello has a BSc (Hons) in Electronic systems engineering from Monterrey Institute of Technology, Monterrey, Mexico (1987) and a PhD in Biomedical systems from Imperial College London (1996).